\documentclass[prl,aps,twocolumn]{revtex4}
\usepackage{graphicx}% Include figure files
\usepackage{dcolumn}% Align table columns on decimal point
\usepackage{bm}% bold math

\begin{document}

%\draft
%\preprint{}

{\bf Comment on "Magnetization Process of Single Molecule Magnets
at Low Temperatures"}

\vspace{0.5cm}

In a recent letter Fernandez et al. \cite{FernRelax} present Monte
Carlo (MC) simulations of relaxing interacting dipoles for various
lattice structures and argue that the short-time behavior of the
magnetization $M(t)$ in an initially depolarized sample ($M(0)=0$)
is lattice structure dependent. They claim that in a simple cubic
(SC) lattice at short times, $\Delta M(t) = \vert M(t) - M(0)
\vert$ follows the usual $\sqrt t$ law, but that in body centered
(BCC) and face centered (FCC) cubic lattices, $\Delta M(t) \sim
t^{0.7}$.

We disagree with this conclusion, and we believe it arises from
finite size effects in the MC simulations in \cite{FernRelax}.

The $\sqrt t$ short-time relaxation is expected when the dipolar
interaction dominates over the hyperfine coupling, ie., when the
spread of dipolar field bias $W_D \gg \xi_o$, where $\xi_o$ is the
range of energies over which the nuclear bias fluctuates
\cite{PSPRL} (in a completely demagnetized sample $W_D$ is several
times larger than the strength $E_D$ of nearest-neighbour dipolar
interactions). It should persist {\it at least} until $t \sim
(W_D/\xi_o) \tau_o$, where $\tau_o$ is a microscopic single
molecule tunneling relaxation time. In the very short time limit
the relaxation is linear in $t$. Since these predictions all
depended on the long-range part of the dipolar interaction, they
should {\it not} depend on lattice structure.

To show this, we perform MC simulations for BCC, FCC and triclinic
lattices with and without periodic boundary conditions (P.B.C.).
At $t=0$ all spins orientations were random and uncorrelated, with
$M(0)=0$ and $\vec{S}_i = \pm \hat{z}$ ($|\hat{z}|=1$). The
dipolar interaction $V_D(\vec{r}_{ij}) = E_D [\vec{S}_i \vec{S}_j
- 3 (\vec{S}_i \vec{r}_{ij})(\vec{S}_j \vec{r}_{ij})/ r^2_{ij}] /
r^3_{ij}$, and dipole fields and spin configurations were updated
at time intervals $\delta t$ by flipping individual spins with
probability $1 - \exp \{- \delta t \; \tau^{-1}_N(\xi) \; exp(-\xi
/ T) \}$. Here $\tau^{-1}_N(\xi)$ is an incoherent relaxation rate
$\tau^{-1}_N(\xi) = \tau^{-1}_0 e^{- |\xi| / \xi_o}$ (see
\cite{PSPRL}); we put $\tau_o = 1$ and assume contact with a bath
at temperature $T$. The total bias energy $\xi = \xi_{dip} - g_e
\mu_B S H_z$, including the internal dipolar contribution
$\xi_{dip}$.

In the case of BCC and FCC clusters with P.B.C., with dimensions
$16 \times 16 \times 16$ for $E_D=10 \xi_o$ we, as in
\cite{FernRelax}, found $M(t) \sim t^{\sim 0.7}$. However, we also
found that in the FCC cluster (for example) the number of spins in
resonance ($|\xi| \lesssim \xi_o$) is very small, much smaller than
for the corresponding SC cluster. This means that in small FCC
clusters, the MC procedure may not accumulate proper statistics.

To solve this problem one can (i) increase the cluster size, or
(ii) increase the fraction of spins in resonance, by reducing
$E_D / \xi_o$. For a BCC lattice (2 molecules per unit cell) we
applied the first scenario. For a FCC lattice (4 molecules per
unit cell) we applied both, taking $E_D = 2.5 \xi_o$. Since in a
completely demagnetized FCC sample $W_D \sim 10 E_D$, then
$W_D / \xi_o \sim 25$, ie., still $>>1$.

The results (Fig. 1) are clear- for larger samples one finds
$\sqrt{t}$ relaxation if $W_D / \xi_o >> 1$, after the usual initial
linear behaviour. Only the duration of the initial transient and the
subsequent $\sqrt t$ are sensitive to the crystal structure.

\begin{figure}[h]
\centering
\vspace{-1.5cm}
\hspace{0.cm}
\includegraphics[scale=0.4]{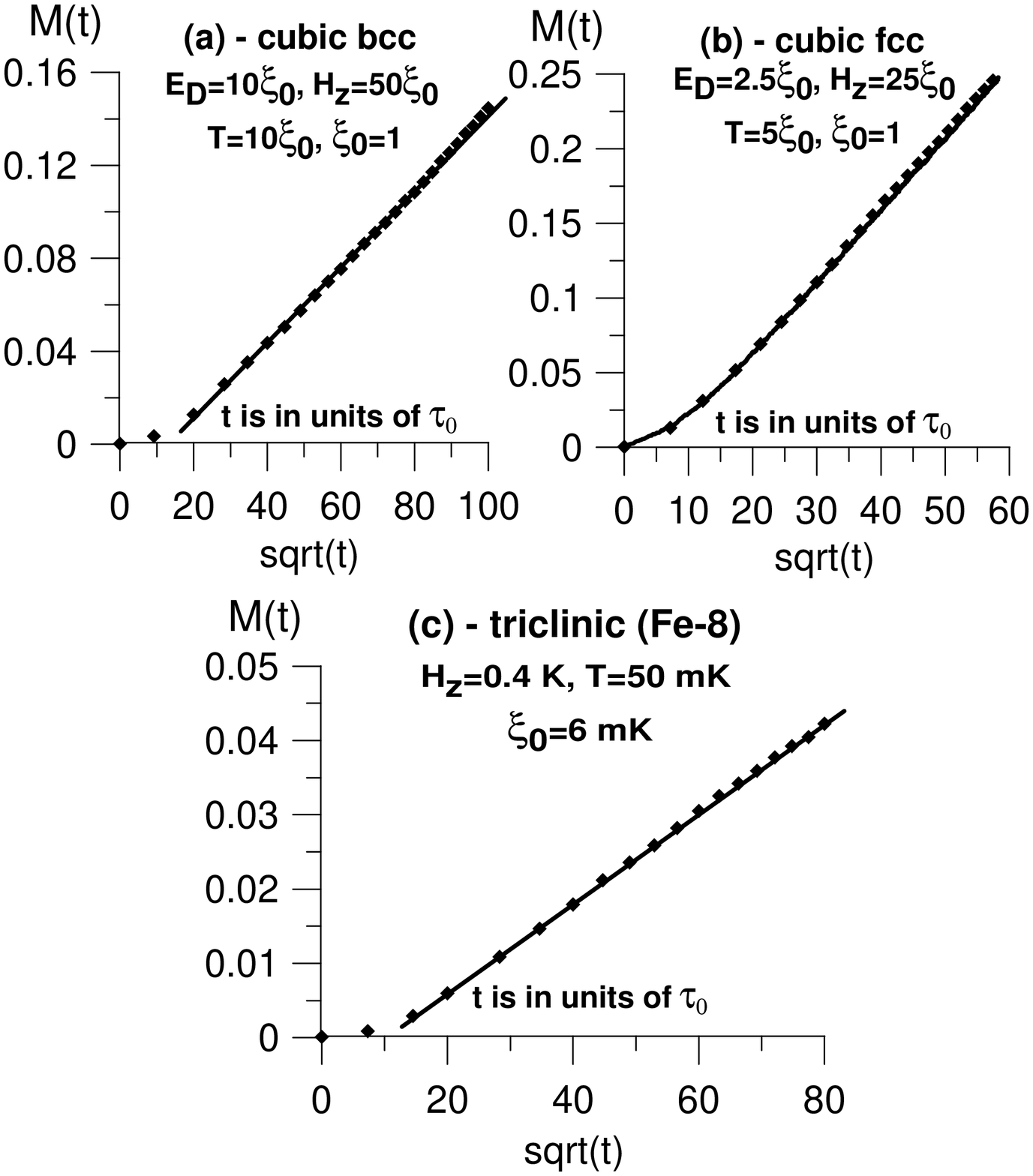}
\vspace{-1.8cm} \caption{$M(t)$ vs $\sqrt t$ for $M(0)=0$. (a) BCC
lattice of size $32 \times 32 \times 32$, with P.B.C.; Diamonds -
the MC result; solid line - the $\sqrt t$ slope of the curve; (b)
FCC lattice, of size $32 \times 32 \times 32$, and P.B.C. (solid
line), and of size $48 \times 48 \times 48$, no P.B.C. (diamonds);
and (c) Triclinic lattice of $60^3$ $Fe-8$ molecules, with no
P.B.C. (Diamonds - the MC result; solid line - the $\sqrt t$ slope
of the curve). Each of the $8$ spin $5/2$ $Fe^{+3}$ ions was
oriented along the easy axis, with lattice positions obtained from
the Cambridge Data base (easy and hard axis orientations were
taken from \cite{BarSes}). All curves were obtained after
averaging over a few hundred random initial configurations of
spins.}
\label{fig:fig1}
\end{figure}

\vspace{0.3cm}

\noindent
\begin{tabular}{l}
I. S. Tupitsyn$^{1}$ and P. C. E. Stamp$^{2}$ \\
$^{1}$ Russian Research Center "Kurchatov Institute", \\
Moscow 123182, Russia. $\;\;\;$ \\
$^{2}$ Physics Department and CIAR, University of British \\
Columbia, 6224 Agricultural Rd., Vancouver BC, \\
Canada, V6T 1Z1.

\end{tabular}

\end{document}